\documentclass[epj]{svjour}
\usepackage{graphics}
\usepackage{amsmath}
\usepackage{amssymb}
\usepackage{mathtools}
\usepackage[bookmarksopen,bookmarksnumbered]{hyperref}
\usepackage{color}
\usepackage{units}
\usepackage{bm}
\usepackage{rotating}
\usepackage{widetext}

\newcommand{\T}[1]{\boldsymbol{#1}_{\text{T}}}
\newcommand{\Tsc}[1]{#1_{\text{T}}}

\newcommand{\bmax}{b_{\rm max}}
\newcommand\mubstar{\mu_{\bstarsc}}

\newcommand{\bigpar}[1]{\left( #1 \right)}
\newcommand{\typeone}{\underline{Type \textrm{I}}: \,}
\newcommand{\typetwo}{\underline{Type \textrm{II}}: \,}

\newcommand\bstar{{\bf b}_*}
\newcommand\bstarsc{b_*}

\DeclareRobustCommand*\diff[2][]{%
   \mathop{
     \mathrm{d}^{#1}
     \mskip-0.2\thinmuskip
    #2}\nolimits
}

\begin{document}
\title{An Overview of Transverse Momentum Dependent Factorization and Evolution}
\author{T.~C.~Rogers
}                     
%
%
\institute{Department of Physics, Old Dominion University, Norfolk, VA 23529, USA \and 
Theory Center, Jefferson Lab, 12000 Jefferson Avenue, Newport News, VA 23606, USA}
\date{September 15, 2015 \; \; \; JLAB-THY-15-2133} 

%
\abstract{
I review TMD factorization and evolution theorems, with an emphasis on the treatment by Collins and originating in the Collins-Soper-Sterman (CSS) formalism.
I summarize basic results while attempting to trace their development over that past several decades. 
} 
\maketitle
%

\section{Introduction}
\label{intro}
I will summarize the basic ideas of the Collins-Soper-Sterman (CSS) 
approach to TMD factorization~\cite{Collins:1981uw,Collins:1981uk,Collins:1981tt,Collins:1982wa,Collins:1984kg}, and the updated version 
in Ref.~\cite[Chapts.~(10,13,14)]{Collins:2011qcdbook}, for formulating transverse momentum dependent factorization.
In this context, ``transverse momentum dependent" (TMD) refers to
QCD treatments of inclusive observables with at least one perturbatively hard scale $Q$, 
and a separate transverse momentum $\Tsc{q}$ that can vary from 
$0$ to order $Q$. It also refers to related objects like TMD parton distribution functions (PDFs) and
TMD fragmentation functions (FFs).
Studies of TMD objects have been driven by a diverse set of motivations that 
include testing perturbative QCD, probing hadronic 
structure, and providing calculations for general particle and nuclear physics experiments.

A trend in TMD physics is that 
initial intuition frequently needs to be revised as quantum field theoretical details come into focus.
I will build on this observation to 
motivate greater general interest in TMD physics as an arena 
for unpacking foundational QCD concepts.

\begin{figure}
  \begin{tabular}{c@{\hspace*{10mm}}c}
\resizebox{0.5\textwidth}{!}{%
  \includegraphics[angle=-90]{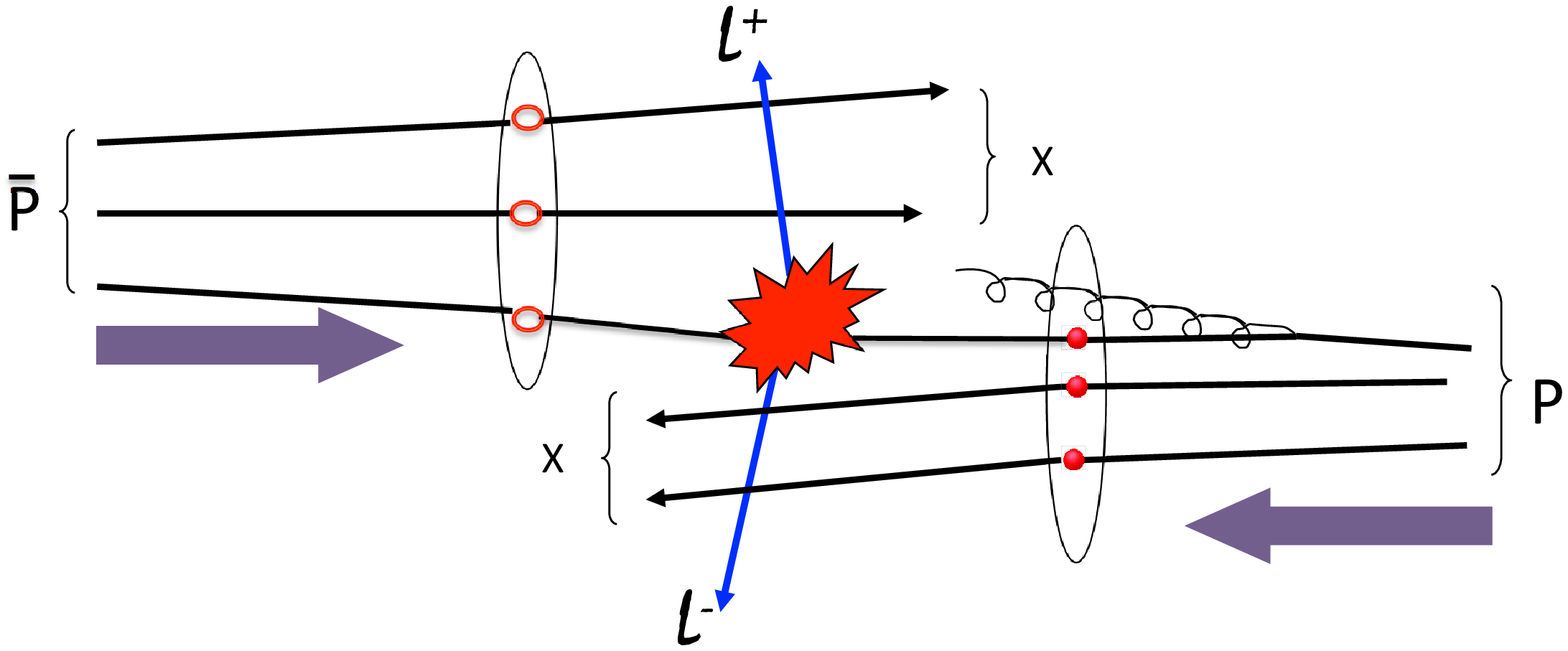}} \\ (a) \\
\resizebox{0.5\textwidth}{!}{%
  \includegraphics[angle=-90]{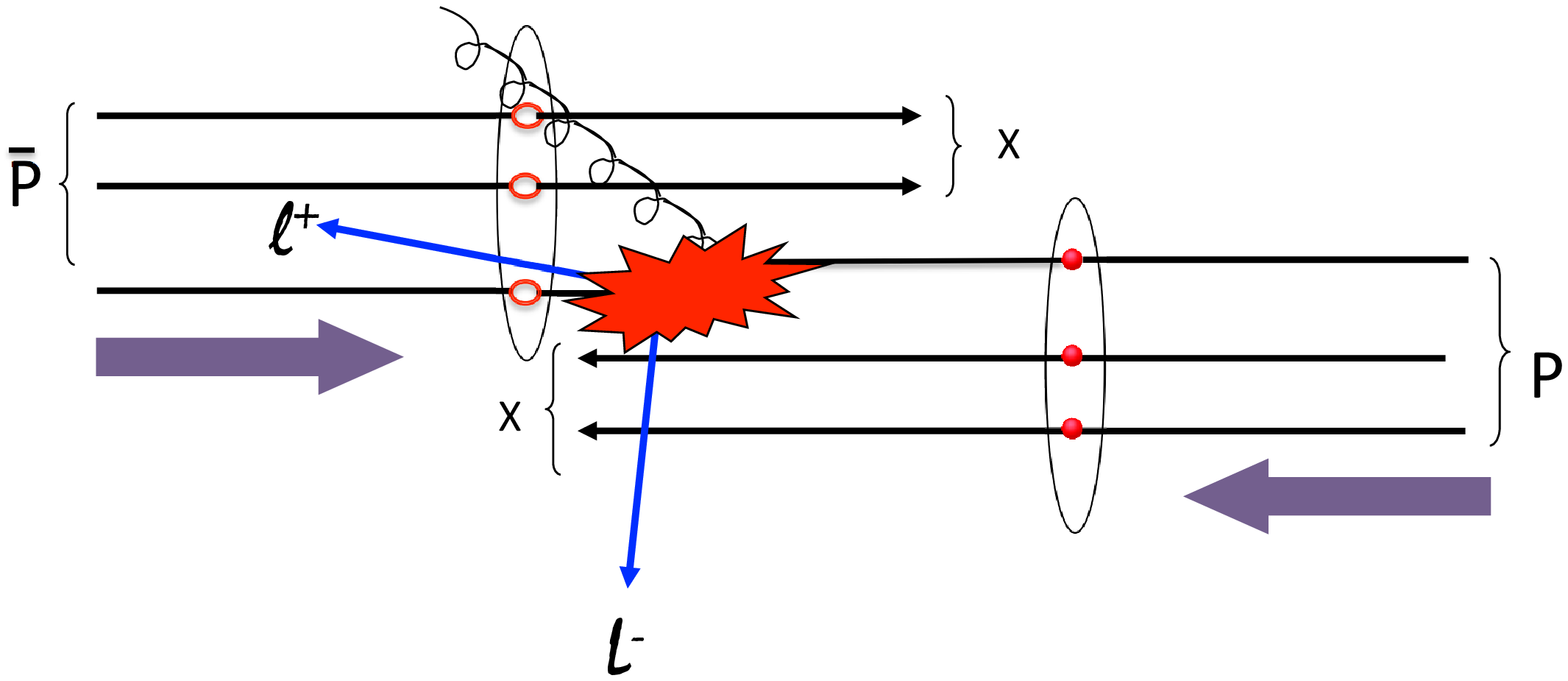}} \\ (b) 
   \\[5mm]
   \end{tabular}
\caption{A cartoon depiction of Drell-Yan scattering with an antiproton entering from the left and a proton 
entering from the right. The explosion is the hard part. In (a), the gluon is taken to be part of the proton wavefunction, making this a 
Type I picture in the language of Ref.~\cite{Feynman:1978dt}.  In (b) the gluon is associated with hard perturbative QCD radiation, making this a type II picture. (Or course, there are many 
more gluons than just the one shown explicitly.)}
\label{fig:1}      
\end{figure}

In its complete form, a TMD factorization theorem should apply to a variety of processes and allow for comparisons 
between them.  But for a concrete discussion, I begin by considering Drell-Yan (DY) scattering. The process is shown 
in cartoon form in Fig.~\ref{fig:1}. In the center of mass frame, a hadron enters from the left with large ``plus" momentum and another enters 
from the right with large ``minus" momentum. An antiquark from one hadron annihilates with a quark from the other, and the resulting 
virtual photon splits into a lepton-antilepton ($l^+ l^-$) pair with total four-momentum $q$. The relevant observable is 
\begin{equation}
 \frac{ \diff{\sigma} }{ \diff[4]{q}\diff{\Omega} } \, , \label{eq:basicprocess}
\end{equation}
where $\diff{\Omega}$ is the phase space of the $l^+ l^-$ pair. 
Dependence on the total transverse momentum $\T{q}$ of the $l^+ l^-$ pair is of special interest at small values and when 
possible correlations to spin dependence are included. Descriptions of Eq.~\eqref{eq:basicprocess} naturally generalize to 
the production of other colorless bosons ($W^{\pm}$, $Z$,Higgs). 

Other processes where TMD factorization applies include semi-inclusive deep inelastic scattering (SIDIS) and $l^+ l^-$ annihilation 
into back-to-back jets or hadrons. In each of these, 
there are two large forward and backward directions analogous to the large positive and negative 
rapidities of the proton and antiproton in Fig.~\ref{fig:1}, and each has a hard scale $Q$ and transverse momentum 
that becomes sensitive to intrinsic transverse momentum when small.

The question of how to describe TMD cross sections like Eq.~\eqref{eq:basicprocess} in QCD
has traditionally been approached from two rather opposite conceptual starting points, which
Feynman, Field and Fox~\cite{Feynman:1978dt} (FFF) identified 
early on, calling them ``Type I" and ``Type II" descriptions. They differentiated between the two pictures as follows:
\begin{description}
\item[\typeone] A description in terms of colliding bound states is taken very literally, with the total transverse momentum 
of the $l^+ l^-$ pair assumed to originate entirely from the nonperturbative intrinsic motion of partons, such as that associated with an incoming  
``wavefunction." (See Fig.~\ref{fig:1}(a).)
\\
\item[\typetwo] The transverse momentum of the $l^+ l^-$ pair is understood to originate from the radiation of 
gluons before and after the collision. In a type II picture, it is assumed that all or most  of such radiation is 
describable using small $\alpha_s$ methods.  (See Fig.~\ref{fig:1}(b).)
\end{description}     
Note the similarity between Fig.~\ref{fig:1} of this paper and Fig.~6 of FFF.
In reference to their Fig.~6, FFF write
\begin{quote}
``There has been much speculation about how much of the dimuon $k_T$  spectra shown in Fig.~7 [DY data from Ref.~\cite{Hom:1976cv,Hom:1976zn,Herb:1977ek,Innes:1977ae}] is due to the 
wave function (Type I) and how much is explained by QCD perturbation calculations (Type II)."
\end{quote}
To some extent, this quote remains true today, though the situation 
is in some ways even more interesting after the advances in QCD theory of the intervening decades. 
   
Although it is a useful starting point for general discussions of TMD physics,
the type I / type II dichotomy is rather artificial, and a sharp distinction 
becomes elusive when one tries to formalize it.  
Figure~6 of FFF already exhibits some of this difficulty. 
Compare, for example, the relatively large intrinsic transverse momentum of 
$\langle \Tsc{k} \rangle \sim 848$~MeV found by FFF with other expectations suggested 
around the same period, such as  the $\langle \Tsc{k} \rangle \sim 300$~MeV proposed in Ref.~\cite{Lam:1977jc} on the basis 
of a parton model.
Also, given what is now understood 
about nonperturbative evolution, a significant fraction of the transverse momentum width is likely actually 
due to nonperturbative radiation that does not fit into either a Type I or Type II category in an obvious way.

In type I oriented approaches of the past, one typically finds discussions of TMD parton models, TMD PDFs, and 
effects from nonperturbative wavefunctions. It is an approach that is used to address problems in hadron structure, such as 
the orbital angular momentum composition of hadrons. See, for example, Ref.~\cite{Liu:2015xha} from this collection for a review.

By contrast, in type II oriented approaches one typically finds discussions 
of fixed high order calculations and/or $\Tsc{q}$-resummation, with nonperturbative transverse momenta only entering in 
the form of small corrections. Applications are to be to high energy physics and very large hard scales, where nonperturbative 
effects tend to get washed out.

One important development of roughly the last decade is a trend toward a convergence of Type I and type II 
oriented approaches into a single TMD formalism. This will be an organizing theme for this review.

In Sect.~\ref{sec:TMDhadstruc} I will expand on the motivations for TMD physics 
that came mainly from the hadron structure perspective, and which has been traditionally seen as 
a more type-I-oriented perspective, while in Sect.~\ref{sec:TMDtype2qcd} I briefly mention the type-II-oriented perspective. 
In Sect.~\ref{sec:TMDpertqcd} I will give an overview of the development of full QCD approaches to TMD factorization.  
In Sect.~\ref{sec:defs}, I will summarize the basic formulas of TMD factorization as they now stand. In Sect.~\ref{sec:solns}, I will discuss 
solutions to the evolution equations, and in Sect.~\ref{sec:summary} I will end with concluding remarks.

\section{Type I Approaches: TMD functions in hadronic structure and nonperturbative physics}
\label{sec:TMDhadstruc}

For describing Type I physics, the parton model is often generalized to include 
an intrinsic transverse momentum for parton distribution functions and fragmentation 
functions -- see, for example, Ref.~\cite[Eq.~(9.13)]{Ellis:1996qj} and the surrounding discussion.
In the DY case, for example, 
one writes
\begin{align}
& \frac{ \diff{\sigma} }{ \diff[4]{q}\diff{\Omega} }  = \sum_{j j'} \mathcal{H}_{j j'} \nonumber \\ & 
\times \int \diff[2]{\T{k}} \, F_{j/A}(x_A,\T{k},S_A) F_{j'/B}(x_B,\T{q} - \T{k},S_B)\, \nonumber \\ 
& \qquad + {\rm p.s.c.} \, . \label{eq:gpmDY}
\end{align}
The basic structure is analogous to collinear factorization, but the usual collinear 
parton distribution functions, with their dependence on longitudinal momentum 
fractions $x_A$ and $x_B$,  are replaced by TMD PDFs with additional dependence on 
the intrinsic transverse momenta (${\bf k}_{T, A} = \T{k}$ and ${\bf k}_{T, B} = \T{q} - \T{k}$).
$F_{f/H}(x_H,\T{k},S_H)$ labels a probability density for finding a parton of flavor $f$ inside a hadron of species $H$. 
The overall hard part is $\mathcal{H}_{j j'}$. It is usually set equal to the zeroth order partonic vertex in a type I approach.
Possible spin dependence is indicated by $S_H$. The ``${\rm p.s.c.}$" means ``power-suppressed corrections."
For processes like SIDIS and $e^+ e^-$ annihilation into back-to-back hadrons, formulas analogous 
to Eq.~\eqref{eq:gpmDY} are needed, but with TMD fragmentation functions. 

From here forward, I will refer to TMD parton distribution functions (TMD PDFs) and 
TMD fragmentation functions (TMD FFs) generically as ``TMDs."

The sensitivity to an intrinsic transverse direction makes TMDs natural objects 
of interest for spin physics because new nonperturbative correlations become possible relative to the collinear leading twist case.
As early as 1978, Cahn used TMD PDFs to describe azimuthal asymmetries in SIDIS~\cite{Cahn:1978se}.
Intrinsic transverse momentum can become correlated with various components of spin and induce asymmetries in the cross section.
In 1990, Sivers proposed a now famous TMD mechanism~\cite{Sivers:1990cc} for explaining the large transverse single spin asymmetries 
in experiments like~\cite{Klem:1976ui,Dragoset:1978gg,Antille:1980th,Apokin:1990ik,Saroff:1989gn,Adams:1991rw,Adams:1991cs}.\footnote{See Ref.~\cite{Aidala:2012mv,Chen:2012taa} for 
some experimental overviews of TMD physics.} 
In the polarized TMD PDF called the Sivers function, the transverse momentum of an unpolarized 
quark becomes correlated with the transverse spin of its parent hadron. 

If one takes a basic number density operator very literally when defining the TMD PDFs, then it appears that time-reversal and parity 
invariance (TP) in QCD requires the Sivers function to vanish. This was pointed out by Collins in Ref.~\cite{Collins:1993kk}, where 
he also proposed an alternative TMD mechanism for generating asymmetries based on fragmentation in the final state. In the ``Collins 
effect," the transverse spins of the quarks involved in the 
hard collision become correlated and remain entangled over long time and distance scales. This results in an 
azimuthal asymmetry in the distribution of unpolarized hadrons. The TMD called the Collins fragmentation function describes the 
likelihood that a transversely polarized quark fragments into a particular unpolarized hadron with some small transverse momentum 
relative to the parent quark.

Over the next decade, the role of TMD functions in describing spin effects grew more prominent in the literature.
In 1995/1996, Mulders and Tangerman classified the leading power azimuthal and spin dependencies  allowed in TMD functions 
by parity and rotation invariance~\cite{Tangerman:1994eh,Mulders:1996dh}.
The ``Boer-Mulders" TMD PDF~\cite{Boer:1997nt} is one particular TMD function that 
illustrates the interesting fact that transverse spin and momentum correlations 
can be important at leading power even in unpolarized scattering.

If the hard scales under consideration cover only a small range, then a reasonable strategy for phenomenology is 
to maintain a purely parton model description, and for a long time this was the most common approach in most applications 
of TMD functions to spin physics and hadron structure. (Reference~\cite{Anselmino:1999pw} 
is an example of this type of application to phenomenology.)
Usually, the TMD is written as a collinear function with a modulating factor for TMD dependence:
\begin{equation}
\label{eq:pmdef1}
F_{q/P}(x,k_T) = f_{q/P}(x) {\rm \Theta}(k_T) \,  ,
\end{equation}
and then requiring 
\begin{equation}
\label{eq:pmcond}
\int d^2 {\bf k}_T \, F_{q/P}(x,k_T) = f_{q/P}(x) \, .
\end{equation}
The modulating factor ${\rm \Theta}(k_T)$ is usually taken to be Gaussian and $x$ and $z$ independent.
See, for example, Ref.~\cite[Eq.~(9.14)]{Ellis:1996qj}. 

While the role of nonperturbative transverse momentum was becoming less of a focus 
in many of the applications of $\Tsc{q}$-resummation to high energy physics, efforts like those summarized in 
the last few paragraphs, particularly when applied to spin physics, focused on intrinsic transverse momentum as a way 
to study non-trivial aspects of fundamental QCD.

However, the work discussed in this section so far was mostly done in the context 
of a TMD parton model or parton-model-like description. 
The situation becomes very interesting when going beyond a TMD parton model picture.
In incorporating perturbative QCD, one expects the TMD functions to acquire scale dependence 
through renormalization group (RG) dependence, analogously to collinear PDFs. Perturbative QCD predicts 
a large transverse momentum dependence that is power-like rather than Gaussian. In 1991, Chay, Ellis and Stirling used 
TMD PDFs to describe azimuthal asymmetries in SIDIS, with a matching to perturbative behavior at 
large transverse momentum~\cite{Chay:1991nh}. A more recent discussion of the matching of large and small transverse momentum 
regions in TMD functions is in Ref.~\cite{Bacchetta:2008xw}.

Much of the work of the full QCD approach began very early, but was 
not commonly incorporated into type-I non-perturbative 
physics studies like those discussed above until relatively recently. 
This will be discussed in more detail in Sect.~\ref{sec:TMDpertqcd}.

\section{Type II physics and collinear factorization}
\label{sec:TMDtype2qcd}

In traditions that approach observables like Eq.~\eqref{eq:basicprocess} from a more Type-II-like perspective, 
one typically starts from calculations of large transverse momentum in perturbative QCD, and attempts to 
extend the description to smaller $q_T$. One works entirely in collinear factorization so that the only nonperturbative 
objects that appear are the collinear PDFs and FFs.
One example is transverse momentum resummation~\cite{Dokshitzer:1978hw,Dokshitzer:1978dr}, 
which incorporates large logarithms of $\Tsc{q}/Q$ to all orders in $\alpha_s$.
These approaches typically assume $\Lambda_{\rm QCD} \ll \Tsc{q} \ll Q$ and ignore intrinsic nonperturbative 
transverse momentum effects, or at least refrain from accounting for them with detailed QCD considerations.
In certain practical circumstances, such as at very high energies and large hard scales, these approaches may be 
sufficient, because sensitivity to nonperturbative transverse momentum become suppressed in the limit of infinite $Q$, even 
down to small $\Tsc{q}$~\cite{Parisi:1979se}. The advantage is that calculations can be done without needing nonperturbative 
input that is often unknown or poorly constrained. However, working from a purely type II approach means that one 
cannot directly connect results to lower $Q$ measurements where nonperturbatuve transverse momentum 
definitely becomes important, and one also abandons the study of 
nonperturbative transverse structure itself via the extraction of TMD functions. 
Of course, in a full QCD treatment resummation-like results should emerge naturally in the large $Q$ limit.

For more discussion of $\Tsc{q}$-resummation, see for example Ref.~\cite[Chapt.~9.0]{Ellis:1996qj} and Ref.~\cite[Chapt.~6]{Greiner:2002ui} and references therein.
See also the cautionary remarks regarding $\Tsc{q}$-resummation techniques in Ref.~\cite[Sec.~13.13.5]{Collins:2011qcdbook}.

\section{TMDs and QCD}
\label{sec:TMDpertqcd}

Notions of intrinsic transverse momentum and TMD functions appeared early on in considerations of full QCD~\cite{Soper:1979fq,Collins:1980ui}.  
It is clear from the FFF discussions that a complete QCD formalism, like a TMD factorization formalism, would involve combination of type I and type II physics.
In fact, the CSS formalism is a TMD factorization, though the way it was originally presented and subsequently applied possibly discouraged its rapid adoption 
in areas like spin physics. An early exception to the tendency to neglect TMD evolution 
in hadron structure phenomenology are the papers of Boer~\cite{Boer:2001he,Boer:2008fr}.  
These are the first cases I know of where CSS-style evolution is applied directly to the phenomenology of single-spin 
asymmetries and azimuthal asymmetries directly identified with TMDs in the type-I sense of Sect.~\ref{sec:TMDhadstruc}. See also Refs.~\cite{Ji:2004wu,Ji:2004xq}.
CSS style treatments similar to Refs.~\cite{Meng:1991da,Meng:1995yn,Nadolsky:1999kb,Nadolsky:2000ky} for unpolarized 
SIDIS were extended to the polarized case in Refs.~\cite{Koike:2006fn}.

An extra complication with TMD parton model approaches is that definitions that use the naive number density operator contain 
extra ``light-cone" divergences. 
The light-cone divergences remain if light-like Wilson lines are used to enforce gauge invariance in TMD definitions, 
even if infrared and ultraviolet divergences are regulated. 
Unlike the normal infrared divergences, which signal the onset of genuine nonperturbative 
physical phenomena in the region of soft physics, the light-cone divergences are artifacts of 
the approximations that separate the cross section into different factors for widely separated regions of rapidity.  
(They are partly artifacts of ignoring the role of soft gluons.)
Light-cone divergences describe gluons with infinite rapidity in the direction \emph{opposite} that of the parent hadron. 
For the approximations to be consistent with factorization, therefore, 
the light-cone divergences need to be regulated and dealt with in some way. 

In 1981, Collins and Soper (CS) introduced operator definitions for TMD PDFs and TMD FFs~\cite[Eqs.~(2.1) and~(4.9)]{Collins:1981uw}, 
and these definitions remain adequate for many purposes.\footnote{The term ``decay function" was
used in Refs.~\cite{Collins:1981uw,Collins:1981uk} rather than ``fragmentation function."} 
The TMD FFs from Ref.~\cite{Collins:1981uw} were used in the derivation of the CS equation  
in Ref.~\cite{Collins:1981uk} -- see \cite[Eq.~(3.8)]{Collins:1981uk}.  
Light-cone divergences were handled by defining TMDs in a non-light-like axial gauge. 
Early discussions of light-cone divergences and the axial-gauge-method of dealing with them 
are discussed in the pioneering work of Refs.~\cite{Soper:1979fq,Ralston:1980pp}. 
(These papers also contain useful references for many of the now standard techniques, such 
as the use of non-light-like axial gauges.)
In particular, Ref.~\cite{Soper:1979fq,Ralston:1980pp} showed how TMD parton correlation functions 
acquire dependence on an auxiliary scale $\zeta = (2 P \cdot n)^2 / (-n^2)$, where $P$ is the proton four-momentum and 
$n$ is a non-light-like gauge fixing vector $n^2 \neq 0$.  
The dependence on $\zeta$ gives rise to logarithmic scaling violations in a complete cross section.  
For the Sudakov  
form factor, a derivation of the corresponding evolution in perturbative QCD was given in Ref.~\cite{Collins:1980ih}, and for the CS 
TMD functions in Refs.~\cite{Collins:1981uw,Collins:1981uk}. 
An analysis of TMD PDFs in structure functions is given in Ref.~\cite{Ralston:1979ys}.
The $\zeta$ scale, with its connection to the gauge fixing vector $n$, can be 
thought of as a cutoff on light-cone divergences.
The CS equation gives the evolution with respect to the direction of the gauge fixing vector $n$
and restores predictive power which would otherwise be lost by having an extra parameter.\footnote{Early work such as 
Refs~\cite{Ralston:1980pp,Collins:1981uw,Collins:1981uk,Collins:1984kg} did not use the 
terminology ``TMD," which became common only later.}  

The TMDs, defined in a non-light-like 
axial gauge as in Refs.~\cite{Collins:1981uw,Collins:1981uk}, use the same auxiliary scales $\zeta$ 
as in Ref.~\cite{Ralston:1980pp}. Since they make explicit use of the gauge fixing vector in treating light-cone divergences, the early 
definitions of the TMDs in Refs.~\cite{Collins:1981uw,Collins:1981uk} were not gauge invariant. 
Treatments in Ref.~\cite{Collins:1981tt,Collins:1982wa} did propose 
gauge invariant definitions for the TMD PDFs, using non-light like Wilson 
lines to regulate light-cone divergences, and similar procedures have come to be preferred.
However, the CSS formalism for hadron-hadron scattering, as it was presented in Ref.~\cite{Collins:1984kg}, was based on 
the earlier derivation of TMD factorization for $e^+ e^-$ annihilation into 
back-to-back jets with the non-light-like axial gauge definitions of Ref.~\cite{Collins:1981uw}. 
The original definitions of the TMD functions are also modified in Ref.~\cite{Collins:1984kg} relative to those of 
Refs.~\cite{Collins:1981uw,Collins:1981uk} such that the overall hard factor is unity and there is no 
explicit $U(b)$ in the factorization formula.\footnote{$U(b)$ is related to the well-known soft factor, often called $S(b)$.}
(See the footnote at Eq.~(3.3) of Ref.~\cite{Collins:1984kg} and the discussion that begins with that equation.)

In most derivations of factorization, especially when initial state hadrons are involved, determining a method for dealing with   
gluons in the ``Glauber" region is a major step toward the ultimate factorization, and it is the source of many 
of the subtleties that affect the TMD definitions. 
The Glauber region describes gluons whose longitudinal momentum components vanish while the the 
transverse components remain small (say $\sim \Lambda_{\rm QCD}$) but fixed. The approximations that would normally allow one to apply Ward identities 
and eikonalize soft and collinear gluons fail in the Glauber region. So Glauber gluons threaten to spoil factorization~\cite{Bodwin:1981fv}.\footnote{An analogy can 
be made between Glauber gluon exchanges at the partonic level and multiple nucleon interactions in a Glauber model of nucleus-nucleus scattering.} 
Therefore, any derivation of factorization (either collinear or TMD) must show that Glauber region contributions either cancel in a sum over all graphs, or 
are avoided by contour deformations in the integrals over gluon momenta. 
Observables that involve collisions between two hadrons are especially 
challenging because gluon exchanges between parton spectators are ``pinched" in the Glauber region, blocking  
straightforward contour deformations. For inclusive DY, the Glauber region contributions can be shown to cancel for 
spectator-spectator interactions~\cite{Collins:1985ue,Collins:1981tt,Collins:1982wa,Lindsay:1982gd,Bodwin:1984hc}.
See also~\cite{Collins:1998ps} and~\cite[Chapts.14.3-14.5]{Collins:2011qcdbook} for more recent pedagogical overviews.\footnote{Reference~\cite[pg.~12]{Collins:1981tt} notes an interesting 
similarity between Glauber cancelations and the AGK unitarity cancelations~\cite{Abramovsky:1973fm} in Regge theory.} 

In the TMD case, many of the steps for showing a cancelation of Glauber region effects, especially in spectator-spectator 
interactions, carry over from the inclusive collinear case. CSS used this in Ref.~\cite{Collins:1984kg} 
to carry TMD factorization results originally obtained for $l^+ l^-$ annihilation over 
to the DY case.
  
The details of how  
Glauber regions are avoided by contour deformations are closely connected to establishing good TMD definitions~\cite{Collins:2004nx}. 
The corresponding subtleties are associated with many of the nonintuitive aspects of 
TMD factorization (see the discussion of the Sivers effect below). 

The original CS definitions in Refs.~\cite{Collins:1981uw,Collins:1981uk}, and subsequent definitions 
in Ref.~\cite{Collins:1984kg} that are based on them, are sufficient for capturing much of the physics needed to 
set up basic factorization theorems.
The CSS presentation is the starting point for successful 
applications to phenomenology, especially in unpolarized Drell-Yan-type processes and $e^+ e^-$ annihilation into back-to-back 
hadrons and jets. It was extended to the SIDIS case in Refs.~\cite{Meng:1991da,Meng:1995yn}.
The CSS formalism in this or roughly similar forms has been widely applied. 
See, for example, Refs.~\cite{Ladinsky:1993zn,Balazs:1995nz,Balazs:1997xd,Nadolsky:1999kb,Nadolsky:2000ky,Landry:1999an,Qiu:2000hf,Qiu:2002mu,Fai:2003zc}.
It is often the most convenient way of formulating cross section calculations 
for some practical calculations, especially in contexts where there is comparatively little
sensitivity to, and/or interest in, the precise details of nonperturbative transverse momentum.  
It also clearly displays the underlying simplicity of solutions to the TMD evolution equations, and it makes very explicit 
the matching to collinear correlation functions and collinear factorization in the limits of large $\Tsc{q}$ and $Q$.

However, these early definitions had shortcomings that made them non-ideal for 
confronting some of the issues discussed in Sect.~\ref{sec:TMDhadstruc}. 
Some of the problems are mainly organizational. 
One possibly confusing aspect of the presentation in Ref.~\cite{Collins:1984kg} is 
that the central result -- that the formalism is, first and foremost, a TMD factorization formalism -- appeared 
only later in the paper, in Eqs.~(5.2,5.8).  
The first equation of the paper, which might appear 
to a reader to be the main result,  contains 
no explicit intrinsic nonperturbative transverse coordinate ($\Tsc{b}$) dependence. Perhaps as a result, the CSS formalism is now frequently referred to as a 
resummation. However, it was intended to be more powerful than resummation methods;  the CSS formalism was 
meant to be a true TMD factorization formalism, with
a valid pQCD perturbation expansion of the hard part and renormalization group equations 
for all $b_T$, even in the limit of $b_T \to \infty$ where $\Tsc{b}$-dependence is nonperturbative \textit{(Collins, private communication)}.

If the hard scale is extremely large, the effects of perturbative radiation becomes so dominant 
that all nonperturbative transverse momentum effects are washed out even for $\Tsc{q} \sim 0$~\cite{Parisi:1979se}.  
The process gradually becomes entirely a type II process. In this sense, questions of whether interesting nonperturbative TMD phenomena 
like the Sivers effect are washed out at larger $Q$ are correlated to the question about whether $Q$ is large 
enough that purely perturbative calculations of transverse momentum dependence are reliable to within a desired accuracy.
In Ref.~\cite{Collins:1984kg}, an estimate was given for the scale where the DY cross section becomes completely 
insensitive to nonperturbative transverse momentum dependence, and a value of $Q \approx 10$~PeV was found.  
Given the relative insensitivity to nonperturbative transverse momentum observed at facilities like the Tevatron~\cite{Landry:2002ix,Konychev:2005iy}, it may be that this 
estimate needs to be revised downward. However, even at scales as large as weak boson masses, the nonperturbative contribution 
seems to be important for currently desired levels of accuracy and precision~\cite{Nadolsky:2004vt,Guzzi:2013aja,LopesdeSa:2013zga}.

Even when nonperturbative $\Tsc{b}$-dependence is included, the factorization formulas of Ref.~\cite{Collins:1984kg} do not immediately
resemble TMD parton model formulas like Eq.~\eqref{eq:gpmDY}. 
The connection between TMD functions (including the many subtleties involved in defining those functions)
and the evolved factors used in actual cross sections becomes somewhat indirect. An example of how this can lead to practical 
consequences can be seen in discussions of the Sivers function. 
In 2002, Brodsky, Hwang, and Schmidt (BHS)~\cite{Brodsky:2002cx} used an explicit model 
calculation to demonstrate that final state interactions in SIDIS can give a transverse single spin asymmetry 
at leading power in $Q$.  This was presented as a conflict with factorization itself; they argued that the effect 
cannot be associated with parton densities or fragmentation functions.  
But Collins showed~\cite{Collins:2002kn} that in fact the result is consistent with TMD factorization.  
Instead the BHS calculation effectively demonstrated that a Sivers-like effect is non-vanishing.   
While the TP-invariance argument \cite{Collins:1993kk} mentioned earlier appeared to show that the 
Sivers function vanished, there is a loophole arising from the Wilson lines in a gauge-invariant definition of 
TMD functions.  When applied to those definitions, TP invariance shows that the Sivers function used in the 
Drell-Yan process has the opposite sign to the one in SIDIS, not that it vanishes.
Thus, what might at first seem like a contradiction between factorization and the direct calculations of Ref.~\cite{Brodsky:2002cx} is largely 
due to an unclear connection between TMD definitions, as they had been presented up to that time, and their origins in a factorization derivation. 
The TMD factorization derivation relies on contour deformations that avoid the Glauber region, and the 
way this procedure gets modified to factorize different processes introduces  
a process dependent sign and the associated Wilson line direction. 

Another issue with the factorization as organized in Ref.~\cite{Collins:1984kg} is that perturbatively calculable process dependence was moved 
out of the hard part and into the factors that most resemble the evolved TMD functions; the hard part is simply $1$. This gives a simple arrangement in some sense. But in a 
truly factorized TMD formulation, one expects all fixed order perturbative process dependence to be explicitly factorized.
This was especially clearly pointed out by de Florian and Grazzini in Refs.~\cite{deFlorian:2000pr,deFlorian:2001zd}. 
Catani et al. subsequently developed an approach to organizing non-universal contributions into a 
separate hard factor~\cite{Catani:2000vq,Catani:2009sm,Catani:2012qa,Bozzi:2010xn,Catani:2013tia}.
But with well-defined TMD definitions, including a specification of a renormalization scheme, the hard part  is uniquely determined automatically from Eq.~\eqref{eq:gpmDY}. 

I have listed several issues that highlight the importance of considering TMD definitions in detail and examining their origins in  
a factorization theorem.
The discussions of the last few paragraphs also motivate natural conditions for optimally  
defining TMD PDFs and FFs: a.) They should have definite operator definitions that are separately 
gauge invariant and account for any instances of nonperturbative process dependent signs. b.) 
Apart from the process dependent signs, they should be universal and provide a prescription 
for calculating hard parts. c.) They should combine perturbative and nonperturbative information in a way 
that allows one to simultaneously extract maximum advantage from the universality of nonperturbative parts and from the 
perturbative calculability of small coupling parts.

In fact, the details of how to precisely define the TMDs becomes a guiding question for setting up 
the newer TMD definitions in next section.  
The universality of the TMD functions needs to be modified from the parton model 
in at least two respects: 1.) there is process dependence in the sense of dependence on a hard 
scale $Q$ via evolution (just as in collinear factorization) and 2.) there is dependence on the direction of the Wilson line. 
The second of these is a novel modification of the more familiar universality concept  encountered 
when dealing with collinear PDFs.

The early 2000s saw 
growing attention paid to the issue of precisely defining TMD functions, largely due to gradually increasing  
recognition of the relevance of predictions like the changing Sivers function sign and the inadequacy of light-cone Wilson lines for definitions. See, especially, 
Refs.~\cite{Collins:1999dz,Belitsky:2002sm,Collins:2003fm,Boer:2003cm,Bomhof:2004aw,Ji:2004wu,Ji:2004xq,Bomhof:2006dp,Cherednikov:2008uk,Cherednikov:2007tw,Cherednikov:2008ua,Hautmann:2007uw,Collins:2008ht,Hautmann:2009zzb}. 
Reference~\cite{Collins:2003fm} contains a useful summary of the issues as they stood 
approximately a decade ago. These considerations led to the formulation that will be discussed in the next section.

\section{Basic statement of TMD factorization}
\label{sec:defs}

The organization of TMD factorization theorems and their TMD definitions, as it will be presented in the remaining sections, is 
based on Ref.~\cite{Collins:2011qcdbook}. This formulation is intended to satisfy the criteria listed at the end of the 
last section. 

\subsection{Factorization formulas}

Expressions with TMDs are most easily expressed in transverse coordinate space:
\begin{align}
& F_{j/P}(x,\T{k},S_A;\mu,\zeta_{\rm PDF}) = \nonumber \\
&{} \qquad \frac{1}{(2 \pi)^2} \int \diff[2]{\T{b}} \, e^{i \T{k} \cdot \T{b}} \, \tilde{F}_{j/P}(x,\T{b},S_A;\mu,\zeta_{\rm PDF}) \, , \label{eq:TMDPDFdefFT} \\
&D_{H/j}(z,z \T{k},S_B;\mu,\zeta_{\rm FF}) = \nonumber \\
&{} \qquad \frac{1}{(2 \pi)^2} \int \diff[2]{\T{b}} \, e^{-i \T{k} \cdot \T{b}} \, \tilde{D}_{H/j}(z,\T{b},S_B;\mu,\zeta_{\rm FF}) \, . \label{eq:TMDFFdefFT}
\end{align}
Note the convention to write the left side of Eq.~\eqref{eq:TMDFFdefFT}  as a function of $z \T{k}$ rather than $\T{k}$. The nonperturbative 
behavior associated with small $\Tsc{q}$ is associated with large $\Tsc{b}$ behavior in the coordinate space functions.

TMD factorization theorems are best established theoretically 
for the classic electromagnetic processes of DY, SIDIS and 
the annihilation of $e^+ e^-$ pairs into a back-to-back hadron pairs. The basic statement of TMD factorization for these three 
processes is~\cite{Collins:2011qcdbook}:

\begin{widetext}
\begin{align}
  \frac{ \diff{\sigma} }{ \diff[2]{\T{q} \cdots } }
  ={}&
   \mathcal{H}^{\rm DY}_{j \bar{\jmath}}(\mu/Q;\alpha_s(\mu))
    \int \diff[2]{\T{b}}
    ~ e^{i\T{q}\cdot \T{b} }
    ~ \tilde{F}^{[-]}_{j/A}(x_A,\T{b},S_A;\zeta,\mu) 
    ~ \tilde{F}^{[-]}_{\bar{\jmath}/B}(x_B,\T{b},S_B;Q^4/\zeta,\mu) \; \; + Y_{\rm DY}\, , & \label{eq:DYfact}
 \\
  \frac{ \diff{\sigma} }{ \diff[2]{\T{q} \cdots } }
  ={}&
   \mathcal{H}^{\rm SIDIS}_{j \bar{\jmath}}(\mu/Q;\alpha_s(\mu))
    \int \diff[2]{\T{b}}
    ~ e^{-i\T{q}\cdot \T{b} }
    ~ \tilde{F}^{[+]}_{j/A}(x_A,\T{b},S_A;\zeta,\mu) 
    ~ \tilde{D}_{B/\bar{\jmath}}(z_A,\T{b},S_B;Q^4/\zeta,\mu) \; \; + Y_{\rm SIDIS}\, , & \label{eq:SIDISfact}
\\
  \frac{ \diff{\sigma} }{ \diff[2]{\T{q}\cdots } }
  ={}&
   \mathcal{H}^{\rm e^+e^-}_{j \bar{\jmath}}(\mu/Q;\alpha_s(\mu))
    \int \diff[2]{\T{b}}
    ~ e^{-i\T{q}\cdot \T{b} }
    ~ \tilde{D}_{A/j}(z_A,\T{b},S_A;\zeta,\mu) 
    ~ \tilde{D}_{B/\bar{\jmath}}(z_B,\T{b},S_B;Q^4/\zeta,\mu) \; \; + Y_{\rm e^+e^-} \, .& \label{eq:epemfact}
\end{align}
\end{widetext}
The left side is a cross section differential in at least transverse momentum $\T{q}$, defined in an appropriate reference 
frame, and the ``$\cdots$" represents possible dependence on other kinematic variables like rapidities.
The first term in each equation has a structure like that of a TMD parton model. For example, use Eq.~\eqref{eq:TMDPDFdefFT} to write
\begin{align}
& \int \diff[2]{\T{b}}
    ~ e^{i\T{q}\cdot \T{b} }
    ~ \tilde{F}^{[-]}_{j/A}(x_A,\T{b},S_A;\zeta,\mu) \times \nonumber \\
 & \qquad  \qquad \qquad ~ \times \tilde{F}^{[-]}_{\bar{\jmath}/B}(x_B,\T{b},S_B;Q^4/\zeta,\mu) \nonumber \\
& = \int \diff[2]{{\bf k}_{T,1}} \int \diff[2]{{\bf k}_{T,2}} 
	    ~ F^{[-]}_{j/A}(x_A,{\bf k}_{T,1},S_A;\zeta,\mu) \times \nonumber \\
& \; ~ \times F^{[-]}_{\bar{\jmath}/B}(x_B,{\bf k}_{T,2},S_B;Q^4/\zeta,\mu) \delta^{(2)}(\T{q} - {\bf k}_{T,1} - {\bf k}_{T,2})\, ,
\end{align}
and compare with Eq.~\eqref{eq:gpmDY}. 
In each of Eqs.~\eqref{eq:DYfact}-\eqref{eq:epemfact}, there is a convolution of two TMD functions, each being either a TMD PDF (labeled by $\tilde{F}$) or a 
TMD FF (labeled by $\tilde{D}$). Capital letters are used here to distinguish TMDs from collinear PDFs and FFs.
The first term is called the ``$W$-term." The $W$-terms differ from 
a parton model picture by the presence of an explicit hard factor $H_{j \bar{\jmath}}(\mu/Q;\alpha_s(\mu))$, and 
by the appearance of evolution scales $\zeta$ and $\mu$. The scales are exactly arbitrary, but in applications they should be set to values 
of order $Q^2$ and $Q$ respectively to enable well-behaved perturbative calculations. The ``$[\pm]$" superscripts 
represent process dependent Wilson line directions, following the notation of~\cite{Boer:2003cm}. 
The $[+]$ means future-pointing and the $[-]$ means past-pointing. The $S_A$ and $S_B$ denote possible polarization dependence. From here forward, the power suppressed 
correction term will be assumed implicit.

Transverse momentum dependence can be factored into separate TMD functions only when $\Tsc{q} \ll Q$.
For very large $\Tsc{q}$, a correction is needed, though it is calculable in pure collinear factorization.
The correction is indicated by the last term in each equation, called the ``$Y$-term."

It is probably best to apply the term ``TMD factorization theorem" to the complete
collection of formulas in Eqs.~\eqref{eq:DYfact}-\eqref{eq:epemfact}, including 
the set of universality properties of large distance parts and the $Y$-terms, rather than to any one equation alone.

\subsection{Definitions}

Each TMD function is defined in terms of quark and gluon field operators.  
I will use SIDIS as a reference process for setting up the definitions, with the directions of the incoming proton and outgoing 
hadron defining the large ``$+$" and ``$-$" directions respectively. 
Define space-like directions by the vectors
\begin{equation}
\label{eq:nLLdir}
n_{\rm A}(y_A) = (1,-e^{-2 y_{\rm A}},{\bf 0}_t) \, , \;\;\; n_{\rm B}(y_B) = (-e^{2 y_{\rm B}},1,{\bf 0}_t) \, .
\end{equation}
These approach light-like plus (minus) vectors when $y_A$($y_B$) approach $+(-) \infty$.

To build up TMD definitions, one first needs to define a soft factor.
Define the Wilson line from  $x$ to $\infty$ along $n$ in terms of a bare coupling and the bare gluon field operator:
\begin{equation}
\label{eq:wildef}
W(\infty ,x;n) = P \exp \left[- ig_0 \int_0^\infty d s \; n \cdot A_0^a (x + s n) t^a \right] \, .
\end{equation}
This Wilson line is in the color triplet representation, with $t_a$ being the ${\rm SU}(3)$ generators in the 
fundamental representation. $P$ is a path-ordering operator.
Following \cite[Eq.~(13.39)]{Collins:2011qcdbook}, the soft factor is the 
vacuum expectation value of a Wilson loop:

\begin{widetext}
\begin{align}
\label{eq:soft}
& \tilde{S}_{(0)}(\T{b};y_a,y_b) \nonumber \\
{}&= \frac{1}{N_C} \langle 0 |W(\T{b}/2,\infty;n_B(y_b))^{\dagger}_{\gamma \alpha} \, W(\T{b}/2,\infty;n_A(y_a))_{\alpha \delta} 
W(-\T{b}/2,\infty;n_B(y_b))_{\beta \gamma} W(-\T{b}/2,\infty;n_A(y_a))^{\dagger}_{\delta \beta} | 0 \rangle_{\rm No \; S.I.}.
\end{align}
\end{widetext} 
The ``${\rm No \; S.I.}$" means Wilson line self-interactions are temporarily excluded, in addition to 
interactions with transverse Wilson lines at light-cone infinity. The Greek letters are color triplet indices. 
An analogous soft factor can be defined in an octet representation. 
The soft factor is designed mainly to describe QCD radiation of gluons that are both nearly on-shell and at central rapidities.

For the TMDs, one would like functions that are reminiscent of number the densities with light-like Wilson lines, but these 
suffer from light-cone divergences as previously discussed. To regulate them while maintaining gauge invariance, 
one starts by defining TMDs with non-light-like Wilson lines.  These are called the ``unsubtracted" definitions.
For quark TMD PDFs and FFs, respectively, they are

\begin{widetext}
\begin{align}
 \tilde{F}_{f/P}^{{\rm unsub},[+]}&(x,\T{b},S;n_B(y_b)) 
= {\rm Tr}_{C,D}  \int \frac{\diff{w^{-}}}{2 \pi} e^{-i x P^+ w^-} \times 
\nonumber \\
  & \times \langle P,S | \bar{\psi}_f \bigpar{\frac{w}{2}} W\bigpar{\frac{w}{2},\infty;n_B(y_b)}^\dagger 
\frac{\gamma^+}{2} W\bigpar{-\frac{w}{2},\infty;n_B(y_b)} \psi_f\bigpar{-\frac{w}{2}} | P,S \rangle_{\rm No \; S.I.} \, , \label{eq:PDFunsub}  \\
 \tilde{D}_{H/f}^{\rm unsub}&(z,\T{b},S;n_A(y_a)) 
=  \sum_X \, \frac{1}{4 z N_{C,f}} {\rm Tr}_{C,D} 
 \int \frac{\diff{w^{+}}}{2 \pi} e^{i k^- w^+} \times 
\nonumber \\
 & \times  \langle 0 | \gamma^- \, W\bigpar{\frac{w}{2},\infty;n_A(y_a)}
\psi_f \bigpar{\frac{w}{2}} | H,S, X \rangle \langle H,S, X  | \bar{\psi}_f\bigpar{-\frac{w}{2}} W\bigpar{-\frac{w}{2},\infty;n_A(y_a)}^\dagger | 0 \rangle_{\rm No \; S.I.} \, . \label{eq:FFunsub}
\end{align}
\end{widetext}
(See~\cite[Eqs.~(13.108,13.41)]{Collins:2011qcdbook} and associated discussions.) 
The definitions here do not use bare fields because the renormalization factors will be included in the final subtracted definition. The ${\rm Tr}_{C,D}$ denotes 
traces over color and Dirac indices. The definition for $\tilde{F}_{f/P}^{{\rm unsub},[-]}(x,\T{b},S;n_B(y_b))$ is exactly the same as in Eq.~\eqref{eq:PDFunsub} but with the main
Wilson lines past pointing rather than future pointing.

The final definitions to be used in Eqs.~\eqref{eq:DYfact}-\eqref{eq:epemfact} are defined with the 
soft factors included, ordinary renormalization, and in the limits of infinite rapidities for the main Wilson lines:
\begin{align}
 & \tilde{F}_{f/P}^{[+]}(x,\T{b};\mu,\zeta_{\rm PDF}) \nonumber \\ &{}=  \lim_{ \substack{ y_A \to +\infty \\ y_B \to -\infty }} F_{f/P}^{{\rm unsub},[+]}(x,\T{b};n_B(y_B)) \times \nonumber \\ &{} \qquad \times
 	\sqrt{\frac{\tilde{S}_{(0)}(\T{b};y_A,y_s)}{\tilde{S}_{(0)}(\T{b};y_s,y_B) \tilde{S}_{(0)}(\T{b};y_A,y_B)}} \times \nonumber \\ 
	   &{} \qquad \; \times {\rm UV \, ren} \, , \label{eq:PDFfinaldef} \\
	   \nonumber \\
 & \tilde{D}_{H/f}(z,\T{b};\mu,\zeta_{\rm FF}) \nonumber \\ &{}=  \lim_{\substack{ y_A \to +\infty \\ y_B \to -\infty }} \tilde{D}_{H/f}^{\rm unsub}(z,\T{b};n_A(y_A))  \times \nonumber \\ &{} \qquad \times
 	\sqrt{\frac{\tilde{S}_{(0)}(\T{b};y_s,y_B)}{\tilde{S}_{(0)}(\T{b};y_A,y_s) \tilde{S}_{(0)}(\T{b};y_A,y_B)}}  \times \nonumber \\ 
	   &{} \qquad \; \times {\rm UV \, ren} \, .  \label{eq:FFfinaldef} 
\end{align}
(See~\cite[Eqs.~(13.106,13.42)]{Collins:2011qcdbook} and associated discussions.) 
The soft rapidity $y_s$ now regulates light-cone divergences.
The factor ``${\rm UV \, ren}$" is an instruction to apply UV renoramlization and remove the UV regulator after the limits of $y_A (y_B) \to +(-) \infty$ have been taken.
In Eqs.~(\ref{eq:PDFfinaldef},\ref{eq:FFfinaldef}), Wilson line self-energies and interactions with transverse Wilson lines at light-cone infinity can be included now because they cancel 
between the soft factors and the main Wilson line in the unsubtracted TMDs. Thus the ``${\rm No \; S.I.}$" has been removed.

The sensitivity to $y_s$ in each TMD is contained in the auxiliary parameters $\zeta_{\rm PDF}$ 
and $\zeta_{\rm FF}$:
\begin{align}
\zeta_{\rm PDF} = x^2 M_P^2 e^{2(y_P - y_s)} \, , \\
\zeta_{\rm FF} = \frac{M_H^2}{z^2} e^{2(y_s - y_H)} \, .
\end{align}
So, 
\begin{equation}
\zeta_{\rm PDF} \zeta_{\rm FF} = Q^4 \, .
\end{equation}
The $\zeta$-parameters carry the memory of the need to regulate Wilson lines to define 
separate TMD functions.

For the Drell-Yan and $e^+ e^-$ processes in Eqs.(\ref{eq:DYfact},\ref{eq:epemfact}), 
the TMD PDFs (FFs) for hadrons moving with large minus (plus) momentum have the same 
definitions but with the plus and minus directions reversed, and corresponding replacements 
of $n_A(n_B)$ with $n_B(n_A)$.

The notational complexity in Eqs.~(\ref{eq:PDFfinaldef},\ref{eq:FFfinaldef}) maybe disguises an 
important simplicity in these definitions.
Divergences are removed by multiplying unsubtracted TMDs by factors with relatively simple and 
universal properties. This is closely analogous to ordinary ultraviolet (UV) renormalization, where 
bare operators are renormalized by multiplying with renormalization factors.
It is useful to define a notation that emphasizes this analogy.
First write
\begin{align}
Z_{\rm CS}(y_s) \equiv \sqrt{\frac{\tilde{S}_{(0)}(\T{b};+\infty,y_s)}{\tilde{S}_{(0)}(\T{b};y_s,-\infty) \tilde{S}_{(0)}(\T{b};+\infty,-\infty)}}\, , \\
\underline{Z}_{\rm CS}(y_s) \equiv \sqrt{\frac{\tilde{S}_{(0)}(\T{b};y_s,-\infty)}{\tilde{S}_{(0)}(\T{b};+\infty,y_s) \tilde{S}_{(0)}(\T{b};+\infty,-\infty)}} \, .
\end{align}
The infinite plus and minus rapidities in the arguments on the right side should 
be taken to mean that one applies the limits of infinity plus and minus rapidities in combination 
with whatever $Z_{\rm CS}(y_s)$, $\underline{Z}_{\rm CS}(y_s)$ multiply on their left.
Then Eqs.~\eqref{eq:PDFfinaldef}-\eqref{eq:FFfinaldef} are
\begin{align}
 & \tilde{F}^{[+]}_{f/P}(x,\T{b};\mu,\zeta_{\rm PDF}) \nonumber \\ &{}=  F_{f/P}^{{\rm unsub}, [+]}(x,\T{b};n_B(-\infty)) Z_{\rm CS}(y_s) Z_{\rm PDF}(\mu) Z_2(\mu) \, , \label{eq:PDFfinaldef2} \\
 \nonumber \\
 & \tilde{D}_{H/f}(z,\T{b};\mu,\zeta_{\rm FF}) \nonumber \\ &{}= \tilde{D}_{H/f}^{\rm unsub}(z,\T{b};n_A(+\infty)) \underline{Z}_{\rm CS}(y_s) Z_{\rm FF}(\mu) Z_2(\mu) \, ,  \label{eq:FFfinaldef2} 
\end{align}
where the products are defined to include first the limits of $y_A (y_B) \to +(-) \infty$, and then 
the application of UV renormalization with removal of UV regulators. The factors 
$Z_{\rm PDF, \, FF} (\mu)$ are the UV renormalization factors for the PDF and FF. The factor $Z_2 (\mu)$ is the 
ordinary UV field strength renormalization factor. 

Limits associated with factors of $Z$ should be taken in order from left to right;
the order of factors in Eqs.~\eqref{eq:PDFfinaldef2}-\eqref{eq:FFfinaldef2} is important.
The limits of infinite Wilson line rapidities needs to be applied before UV regulators are removed (i.e., before $\epsilon \to 0$ in 
dimensional regularization). See~\cite[Sect.~(10.8.2)]{Collins:2011qcdbook} for a detailed discussion of the non-commuting limits.

Note that all the dependence on 
$y_s$ (or, equivalently, $\zeta$) is in $Z_{\rm CS}(y_s)$.  The UV factors 
and rapidity renormalization factors are independent of the nature of the target or measured hadrons.
The overall cross section is independent of $\mu$ and $y_s$, but separate factors 
acquire $\mu$ and $y_s$ dependence from $Z$-factors.
In this sense, $Z_{\rm CS}(y_s)$
is very much like a generalization of a standard renormalization factor. 

\subsection{Evolution}
\label{sec:evolution}

The ordinary renormalization plus rapidity evolution in Eqs.~\eqref{eq:PDFfinaldef2}-\eqref{eq:FFfinaldef2} 
gives a system of evolution equations. (See, for example,~\cite[Eqs.~(13.47,13.49,13.50)]{Collins:2011qcdbook}.) The rapidity evolution equation for the TMD PDF is 
\begin{equation}
\label{eq:CSS.evol}
  \frac{ \partial \ln \tilde{F}^{[\pm]}_{f/P}(x,\Tsc{b}; \mu, \zeta) }
       { \partial \ln \sqrt{\zeta} }
  = 
  \tilde{K}(\Tsc{b};\mu) \, .
\end{equation}
The right side is the CS evolution kernel $\tilde{K}(\Tsc{b};\mu)$, which is 
calculable in perturbation theory at small $\Tsc{b}$ and using $\sim 1 / \Tsc{b}$ as a hard scale. 
It obeys its own RG
equation:
\begin{equation}
\label{eq:RG.K}
  \frac{ \diff{\tilde{K}(\Tsc{b};\mu)} }{ \diff{\ln \mu } }
  = -\gamma_K \left(\alpha_s(\mu)\right) \, .
\end{equation}
The RG equation for the TMD PDF is 
\begin{align}
\label{eq:RG.TMD.pdf}
  & \frac{ \diff{ \ln \tilde{F}^{[\pm]}_{j/P}(x,\Tsc{b};\mu,\zeta) }}
       { \diff{\ln \mu} } \nonumber \\
    & = \gamma_{j,{\rm PDF}}( \alpha_s(\mu); 1 )
      - \frac12 \gamma_K(\alpha_s(\mu)) \ln \frac{ \zeta }{ \mu^2 } \, .
\end{align}
$\gamma_K$ and $\gamma_{j, {\rm PDF}}$ are anomalous dimensions for $\tilde{K}(\Tsc{b};\mu)$ and the TMD PDF respectively.\footnote{A function that is basically 
equivalent to $\tilde{K}$ is called
$-2 D$ in~\cite{GarciaEchevarria:2011rb} and
  $-F_{q\bar{q}}$ in~\cite{Becher:2010tm}.}
At small $\Tsc{b}$, $1/ \Tsc{b}$ becomes a hard scale and the individual 
TMD PDFs can be expanded in a perturbative series in terms of collinear PDFs using an operator product expansion (OPE):
\begin{align}
\label{eq:TMD.OPE}
   \tilde{F}_{j/H}(x,\Tsc{b};\zeta,\mu) 
  = {}& \sum_k \int_{x-}^{1+} \frac{ \diff{\xi} }{ \xi }
       \,\tilde{C}_{j/k}\left( x/\xi,\Tsc{b};\zeta,\mu,\alpha_s(\mu) \right)
\nonumber\\
&\times
        f_{k/H}(\xi;\mu)
~+~ O\left[(m\Tsc{b})^p \right] \, .
\end{align}
The $C_{j/k}$ are Wilson coefficients, and $p > 0$. Equation~\eqref{eq:TMD.OPE} is for an unpolarized 
TMD PDF so I have dropped the $[\pm]$ for the Wilson line direction. TMD PDFs like the Sivers function need to be expanded in 
terms of twist three hard coefficients in the small $\Tsc{b}$ limit.
A set of equations analogous to \eqref{eq:CSS.evol}-\eqref{eq:TMD.OPE} holds for TMD FFs.

The right side of Eq.~\eqref{eq:CSS.evol} is perturbatively calculable 
if $1/\Tsc{b}$ is much larger than $O(\Lambda_{\rm QCD})$ and $\mu$ is fixed to $\sim 1/\Tsc{b}$.
The anomalous dimensions $\gamma_K \left(\alpha_s(\mu)\right)$ and $\gamma_{j \, {\rm PDF}}( \alpha_s(\mu); 1 )$ are perturbatively calculable as long as $\mu$ is much larger than $O(\Lambda_{\rm QCD})$.

One striking difference between TMD evolution and collinear evolution is that Eq.~\eqref{eq:RG.K} implies that, in the limit of large $\Tsc{b}$,  
the evolution itself becomes nonperturbative. Predictive power is maintained because the $\tilde{K}(\Tsc{b};\mu)$ has \emph{strong universality}, 
meaning it is independent not only of the process, but also the species of hadrons involved or any polarizations involved. It is even the same $\tilde{K}(\Tsc{b};\mu)$ 
for TMDs and FFs. It follows from the universality of the $Z_{\rm CS}(y_s)$ from the previous section. 
Testing the strong universality of nonperturbative evolution is an important part of TMD phenomenology. 

\section{Solutions}
\label{sec:solns}
The TMD evolution equations 
only involve products of factors in transverse coordinate space, making  
solutions simple to write. 
In preparation for writing the solutions, one needs several definitions 
associated with the organization of perturbative and nonperturbative parts. 

For small $\Tsc{b}$, the right sides of Eqs.~\eqref{eq:CSS.evol} and~\eqref{eq:TMD.OPE} can 
be calculated entirely in collinear perturbation theory with $1/\Tsc{b}$ acting as the hard scale. The only nonperturbative 
inputs then are the collinear PDFs and FFs. To define what one means by ``large" and ``small" $\Tsc{b}$, one must define 
a cutoff scale $\bmax$. Above $\bmax$, one allows for nonperturbative $\Tsc{b}$-dependence, while below $\bmax$ 
one relies on collinear perturbation theory. A standard procedure is to define a $\bstar(\T{b})$ such that 
\begin{equation}
{\bm b}_*({\bm b}_T) = 
\begin{dcases}
{\bm b}_T & b_T \ll b_{\rm max} \\
{\bm b}_{\rm max} & b_T \gg b_{\rm max} \, . \label{eq:bdef}
\end{dcases}
\end{equation}
A smooth transition function is typically used. One of the most common is~\cite{Collins:1981va}
\begin{equation}
{\bm b}_*(\T{b}) \equiv  \hat{{\bm b}} \sqrt{\frac{b_T^2}{1 + b_T^2 / b_{\rm max}^2}} \, .
\end{equation}
One ultimately evolves to a scale
\begin{equation}
\mubstar \equiv C_1/\bstarsc \,  .
\end{equation}

The perturbative and nonperturbative $\Tsc{b}$-dependence in $\tilde{K}(\Tsc{b};\mu)$ is separated by defining 
\begin{align}
-g_K(\Tsc{b};\bmax)  =&{} -\tilde{K}(\bstarsc;\mu_0)  + \tilde{K}(\Tsc{b};\mu_0) \, , \label{eq:gkdef}
\end{align}
which by construction vanishes like a power at small $\Tsc{b}$. In Eq.~\eqref{eq:gkdef}, the scale $\mu_0$ is arbitrary.

Solving Eqs.~\eqref{eq:PDFfinaldef2}-\eqref{eq:FFfinaldef2} (and the equivalent equations for the FFs) to obtain the 
TMDs at arbitrary scales $\zeta$ and $\mu$ gives:

\begin{widetext}
\begin{eqnarray}
\label{eq:evolvedTMDPDF}
 \tilde{F}_{f/P}(x,{\bf b}_T;\mu,\zeta)
 & = & \stackrel{\rm AA}{\overbrace{\sum_j \int_x^1 \frac{\diff{\xi}}{\xi} {\color{red} \tilde{C}^{\rm PDF}_{f/j}(x/\xi,b_{\ast};\mubstar^2,\mubstar,\alpha_s(\mubstar))} \, {\color{blue} f_{j/P}(\xi,\mubstar)}}} \nonumber \\
& \times & \stackrel{\rm BB}{\overbrace{ \exp \left\{ \ln \frac{\sqrt{\zeta}}{\mubstar} {\color{red} \tilde{K}(b_{\ast};\mubstar)} + 
\int_{\mubstar}^{\mu} \frac{\diff{\mu^\prime}}{\mu^\prime} \left[ {\color{red} \gamma_{f,{\rm PDF}}(\alpha_s(\mu^\prime);1) }
- \ln \frac{\sqrt{\zeta}}{\mu^\prime} {\color{red} \gamma_K(\alpha_s(\mu^\prime)) }\right]\right\}}} \nonumber \\
& \times & 
\stackrel{\rm CC}{\overbrace{\exp \left\{ - {\color{blue} g_{f/P}(x,b_T;\bmax) } - {\color{blue} g_K(b_T;\bmax) } \ln \frac{\sqrt{\zeta}}{Q_0} \right\}}} \, ,
\end{eqnarray}
and 
 \begin{eqnarray}
\label{eq:evolvedTMDFF}
 \tilde{D}_{H/f}(z,{\bf b}_T;\mu,\zeta)
 & = & \stackrel{\rm AA}{\overbrace{\sum_j \int_z^1 \frac{\diff{\xi}}{\xi^3}  {\color{red} \tilde{C}^{\rm FF}_{j/f}(z/\xi,b_{\ast};\mubstar^2,\mubstar,\alpha_s(\mubstar))} {\color{blue} d_{H/j}(\xi,\mubstar)}}} \nonumber \\
& \times & \stackrel{\rm BB}{\overbrace{ \exp \left\{ \ln \frac{\sqrt{\zeta}}{\mubstar} {\color{red} \tilde{K}(b_{\ast};\mubstar)} + 
\int_{\mubstar}^{\mu} \frac{\diff{\mu^\prime}}{\mu^\prime} \left[ {\color{red} \gamma_{f,{\rm FF}}(\alpha_s(\mu^\prime);1) }
- \ln \frac{\sqrt{\zeta}}{\mu^\prime} {\color{red} \gamma_K(\alpha_s(\mu^\prime)) } \right]\right\}}} \nonumber \\
& \times & 
\stackrel{\rm CC}{\overbrace{\exp \left\{ - {\color{blue} g_{H/f}(z,b_T;\bmax)} - {\color{blue} g_K(b_T;\bmax) } \ln \frac{\sqrt{\zeta}}{Q_0} \right\}}} \, .
\end{eqnarray} 
\end{widetext}
In practice, one usually sets $\mu = \sqrt{\zeta} = Q$ to enable perturbative calculations.

I have organized the solutions here into three factors, labeled ``AA," ``BB," and ``CC,"  following the method of Ref.~\cite{Aybat:2011zv} to highlight 
the different components of an evolved TMD and connect type I and type II pictures.\footnote{I use double letters here to distinguish from other common uses of ``A," ``B," and ``C." }

The AA factor is a fixed order calculation in collinear perturbation theory of the small $\Tsc{b}$ dependence.
The BB factor is a perturbative evolution factor for relating scales $\mubstar$ and $\mubstar^2$ to general $\mu$ and $\zeta$. 
The CC factor in the last line includes all nonperturbative transverse coordinate dependence. 
Note the $g_K(b_T;\bmax)$ from Eq.~\eqref{eq:gkdef}. The $g_{f/P}(x,b_T;\bmax)$ and $g_{H/f}(z,b_T;\bmax)$ parametrize 
the transition from the OPE calculation at fixed scale in Eq.~\eqref{eq:TMD.OPE}  to the region where nonperturbative $\Tsc{b}$ dependence 
is included. I have highlighted the roles of perturbative and nonperturbative behavior by making functions 
that are to be calculated entirely in fixed order perturbation theory red while those that include nonperturbative behavior are in blue.
Note in particular that $g_K(b_T;\bmax)$ has no scale dependence, no subscript for $f$, $P$ or $H$, no $x$ or $z$ dependence, and is 
the same function in both Eq.~\eqref{eq:evolvedTMDPDF} and Eq.~\eqref{eq:evolvedTMDFF}. This emphasizes its strong universality, discussed earlier.

The $g_{f/P}(x,b_T;\bmax)$, $g_{H/f}(z,b_T;\bmax)$ and $g_K(b_T;\bmax)$ functions all show 
their explicit dependence on $\bmax$. Since it is an arbitrary cutoff, the overall cross section is 
exactly independent of $\bmax$. If $\bmax$ is varied,  changes in perturbatively calculated parts 
should be compensated by changes in $g_{f/P}(x,b_T;\bmax)$, $g_{H/f}(z,b_T;\bmax)$ and $g_K(b_T;\bmax)$. 
This was exploited in Ref.~\cite{Collins:2014jpa} to improve the 
transition between the dominantly perturbative and dominantly nonperturbative regions of $\Tsc{b}$-dependence.

The solutions in Eqs.~\eqref{eq:evolvedTMDPDF}-\eqref{eq:evolvedTMDFF} are written in a way that maximizes the amount 
of perturbative input, but there are other ways of writing solutions that might be preferred, depending on the context. This was 
discussed recently in Ref.~\cite{Collins:2014jpa}. For example, to match to an exactly parton-model-like picture for some initial 
scale $Q_0$,~\cite[Eq.~(24)]{Collins:2014jpa} may be preferred. If one wishes to organize evolution with respect to center of mass energy $\sqrt{s}$ in 
Drell-Yan, rather than $Q$, then \cite[Eq.~(18)]{Collins:2014jpa} might be preferred.

Equations~\eqref{eq:evolvedTMDPDF}-\eqref{eq:evolvedTMDFF} are written here for unpolarized and azimuthally symmetric functions.
There are similar formulas for other TMDs like the Sivers function~\cite{Aybat:2011ge,Bacchetta:2013pqa}.

\section{Comments}
\label{sec:summary}

I will end with some general remarks about the outlook of TMD applications and work that I believe is still needed.

One important refinement needed for TMD evolution is to include flavor number 
transitions in the full TMD evolution analogous to what is done for collinear PDFs in the ACOT~\cite{Aivazis:1993pi,Collins:1998rz} formalism.
Another complication is with the implementation of $Y$-term corrections. Achieving a smooth matching 
between the $W$-term and the $Y$-term is more complicated in practice than a straightforward implementation 
of the definitions in the factorization derivation~\cite{Boglione:2014oea}. More progress in this area is likely possible with further refinements in the details of  
implementations.

Improvements in nonperturbative theory in treating the TMDs at low $Q$ and low $\Tsc{q}$ 
will be important for phenomenology since the factorization theorems alone provide few detailed 
constraints on the these three dimensional objects. Specific methods will be discussed in other articles in this 
collection. Since TMDs describe inclusive processes, including the radiation of soft hadrons, efforts to constrain them will 
benefit from more detailed pictures of hadronization and fragmentation. An interesting example of fragmentation 
theory applications to TMDs is the use of the string model to describe the Collins mechanism in Ref.~\cite{Artru:1995bh}.
My perspective is that accounting for the constraints of factorization is critical for guiding the formulation of 
general pictures of the underlying physics.

There are by now many other formulations of TMD factorization (or frameworks closely analogous to TMD factorization), and unfortunately it 
is not possible discuss any one of them in detail here. 
An especially active approach in recent years is soft-collinear effective theory (SCET).
There are at least three versions of TMD factorization that start from  
the perspective of SCET~\cite{Mantry:2009qz,Mantry:2010bi,Becher:2010tm,GarciaEchevarria:2011rb}.
Another approach to TMD factorization is that of Ji, Ma, and Yuan~\cite{Ji:2004wu}.

It is likely that insight can be gained by determining if and how different formulations of TMD factorization 
have meaningful differences, or whether they are actually equivalent formulations with 
different notation and/or conventions for intermediate steps. In Ref.~\cite{Collins:2012uy}, a particular version 
of SCET was shown to be equivalent to the TMD factorization approach described in 
Sects.~\ref{sec:TMDpertqcd}-\ref{sec:solns} of this article, at least to one-loop order.
Other issues to consider are the small and large $x$ limits, higher twist corrections, and the relationship 
to exclusive scattering. I refer the reader to other recent reviews such as~\cite{Angeles-Martinez:2015sea,Chen:2015tca} 
for discussions of some of these topics and relevant references.

Detailed theoretical considerations indicate that TMD factorization should break down in some processes where more familiar 
parton model intuition might suggest that it applies~\cite{Collins:2007nk,Collins:2007jp,Rogers:2010dm,Catani:2011st,Forshaw:2012bi,Rogers:2013zha}.   The mechanisms for 
TMD factorization breaking have the potential to produce interesting physical effects themselves,
though more work is needed to determine how to calculate them. For high energies, effects associated 
with TMD factorization breaking are likely calculable in perturbation theory in the form of higher order large logarithms and resummation techniques.

Finally, TMD factorization is also expected to breakdown in certain kinematical limits. For example, 
when target and hadron masses are important, or when the distribution of remnant masses are considered in detail, 
the approximations that give TMD factorization no longer suffice and corrections are needed. 
In such cases, it might be necessary to formulate other forms of factorization.
For example, one might need something more 
like the fully unintegrated factorization advocated in 
Refs.~\cite{Bauer:2007ad,Collins:2007ph,Rogers:2008jk,Jadach:2009gm,Jain:2011iu}, but probably 
more closely analogous to TMD factorization 
as it is formulated in Sect.~\ref{sec:TMDpertqcd}-\ref{sec:solns} of this review.

\begin{acknowledgement}
  I acknowledge many useful conversations with J.~Collins who also provided suggestions on the text.
  I also thank C.~Aidala and R.~Fatemi for helpful comments on the text.
  This work was supported by the DOE contract No. DE-AC05-06OR23177,
  under which Jefferson Science Associates, LLC operates Jefferson Lab. 
\end{acknowledgement}

\bibliographystyle{unsrt}
 \bibliography{tmdreview}

\end{document}